\documentclass[aps,prd,nofootinbib,superscriptaddress,12pt]{revtex4}
\usepackage[hypertex]{hyperref}
\usepackage{graphicx} 
\usepackage{amsbsy}   
\newcommand{\beq}{\begin{eqnarray}}
\newcommand{\eeq}{\end{eqnarray}}

\def\ie{{i.e.},~}
\def\eg{{e.g.},~}
\def\etal{{et al.}~}
\def\ed{\end{document}}
\newcommand{\mz}{\langle{\rm M}_Z\rangle}
\newcommand{\mer}{\dot \rho_Z}
\newcommand{\rsn}{{\rm R}_{\rm SN}}
\newcommand{\msun}{{\rm M}_\odot}
\newcommand{\yr}{{\rm yr}}
\newcommand{\mpc}{{\rm Mpc}}
\newcommand{\sfr}{\dot \rho_*}
\def\lsim{\lower0.6ex\vbox{\hbox{$ \buildrel{\textstyle <}\over{\sim}\ $}}}
\def\gsim{\lower0.6ex\vbox{\hbox{$ \buildrel{\textstyle >}\over{\sim}\ $}}}
\def\SK{Super-K}
\def\kam{KamLAND}
\def\snii{SN\thinspace{$\scriptstyle{\rm II}$}}
\def\ii{thinspace{$\scriptstyle{\rm II}$}~}
\def\OmegaM{\Omega_{\rm M}}
\def\DeltaMsol{\Delta{\rm{m}_{\rm{sol}}^{2}}}
\def\DeltaMatm{\Delta{\rm{m}_{\rm{atm}}^{2}}}
\def\epm{$e^{\pm}\;$}
\def\barnue{\bar{\nu}_{\rm e}}
\def\barnumu{\bar{\nu}_{\mu}}
\def\barnutau{\bar{\nu}_{\tau}}
\def\nue{\nu_{e}}
\def\numu{\nu_{\mu}}
\def\nutau{\nu_{\tau}}
\def\eplus{{\rm e}^{+}}
\def\eminus{{\rm e}^{-}}
\def\Fbare{{\rm F}_{\bar{{\rm e}}}}
\def\Fbare0{{\rm F}_{\bar{{\rm e}}}^{{\rm o}}}
\def\Fbarx0{{\rm F}_{\bar{{\rm x}}}^{{\rm o}}}
\def\Ue1{\vert {\rm U}_{{\rm e}1} \vert}
\def\Ue3{\vert {\rm U}_{{\rm e}3} \vert}
\def\Tnubare{\langle {\rm T}_{\bar{\nu}_{{\rm e}}} \rangle}
\def\Enubare{\langle {\rm E}_{\bar{\nu}_{{\rm e}}} \rangle}
\def\Tnubarmu{\langle {\rm T}_{\bar{\nu}_{\mu}} \rangle}
\def\Tnubartau{\langle {\rm T}_{\bar{\nu}_{\tau}} \rangle}
\def\Enubarmu{\langle {\rm E}_{\bar{\nu}_{\mu}} \rangle}

\begin{document}

\pagestyle{plain}
\title{Mini Z$'$ Burst from Relic Supernova Neutrinos and Late
  Neutrino Masses}

\author{Haim Goldberg}
\affiliation{Department of Physics,
  Northeastern University,
Boston, MA 02115}

\author{Gilad Perez}
\affiliation{Theoretical Physics Group,
Ernest Orlando Lawrence Berkeley National Laboratory,
University of California, Berkeley, CA 94720}

\author{Ina Sarcevic}
\affiliation{Department of Physics,
   University of Arizona,  Tucson AZ 85721}

\date{\today}

\begin{abstract}
   In models in which neutrinos are light, due to a low scale of symmetry
 breaking,
 additional light bosons are generically present.
We show that the interaction between diffuse supernova relic neutrinos
(SRN) and the cosmic background neutrinos, via exchange of these light
scalars, can result in a dramatic change
 of the supernova (SN) neutrinos flux.  Measurement of this effect with current or future
experiments can provide a
 spectacular direct evidence for the low scale models.
We demonstrate how the observation of neutrinos from SN1987A
constrains the symmetry breaking scale of the above models.
We also discuss how current and future experiments
may confirm or further constrain the above models,
either by detecting the ``accumulative resonance'' that diffuse SRN go through
or via a large suppression of the flux of  neutrinos from nearby
$\lsim {\cal O}$ (Mpc) SN bursts.
\end{abstract} \pacs{Who cares?} \maketitle

\section{Introduction}
\label{intro}

The observation of neutrino flavor changing from
solar~\cite{sol,SKsol02,SNO123},
atmospheric~\cite{SKatmnu04} and
terrestrial~\cite{Kam96,K2K} neutrino data has provided firm
evidence for neutrino flavor conversion. The recent new Super-Kamiokande (SK) data
on the $L/E$-dependence of atmospheric neutrino events
\cite{SKoc}, $L$ being the distance traveled by neutrinos of
energy $E$, and the new spectrum data from terrestrial
experiments~\cite{KL766,K2Knu04}, has yielded for the first time
evidence of the expected oscillatory behavior. This strongly
favors non-vanishing sub-eV neutrino masses. These
outstanding developments on the experimental side of neutrino
physics have placed a distinct burden on theorists---to
understand what is the origin of these tiny neutrino masses.

The most elegant and popular solution to this puzzle is the
seesaw mechanism~\cite{seesaw}.
In this
scenario one assumes that lepton number is violated at some high
scale $\Lambda_{\rm L}$ in the form of right-handed neutrino
Majorana masses. This induces, at a lower scale, an effective
operator of the form ${\cal O}(1)\times{(LH)^2/ \Lambda_{\rm
L}}\,,$ where $L$ denotes a lepton doublet and $H$ the Higgs
field. The oscillation data then imply that $\Lambda_{\rm L}\sim
10^{14}\,{\rm GeV}\,.$ While the seesaw mechanism is very
appealing from the theoretical side, it is unlikely to be subject
to direct experimental test sometime in the near future.
An additional virtue of the seesaw mechanism is that it can naturally
provide a platform for  generating the  observed baryon asymmetry of the
universe through leptogenesis~\cite{Lep}.
Introduction of such a high scale, however, requires a mechanism
for electroweak-symmetry-breaking-scale stabilization which
typically leads to various
moduli/gravitino problems in the context of cosmology.
Thus it
is important to explore alternate origins for neutrino masses.

 One such alternative is the late neutrino mass
framework that induces small neutrino masses due to a low scale of
symmetry breaking~\cite{Glob1}.
This idea
points to a completely different understanding for the origin of
neutrino masses. The neutrino masses are protected by some flavor
symmetry different from the one related to the charged fermion
masses. When this symmetry is (say spontaneously) broken by a set
of flavor symmetry breaking vevs, $f$, of fields $\phi$, the
neutrinos acquire masses from $\left( \frac{\phi}{M_F} \right)^n L
N H$ for Dirac neutrinos, or $\left( \frac{\phi}{M_F} \right)^n L
N H + M_R NN$ for Majorana neutrinos, where $N$ denotes a right
handed neutrino, $L,H$ stand for the SM lepton doublet and Higgs
fields respectively
and $M_F$ is a scale in which flavor dynamics takes place~\cite{HMP}.
We want to stress that these
textures do not depend on the details of the symmetry mechanisms,
whether global~\cite{Glob1} or gauge~\cite{Gauge}. Furthermore a
similar scenario can be realized via strongly coupled dynamics
where the compositeness scale is given by $f$~\cite{AG,Take}.

With this alternate scenario, it is then of immediate import to
delimit the allowed range for the symmetry breaking scale, $f$ at
which new physics (NP) appears. Since the principal consequences
of the symmetry breaking are neutrino masses and the relevant new
degrees of freedom couple only to neutrinos, direct experimental
limits on the parameters of this model are unlikely to be
attained.  In fact, the strongest limits on $f$ come from
cosmology and astrophysics rather than from laboratory
data~\cite{Glob1}. As will be discussed shortly, there are
generically associated with this mechanism some extra light
degrees of freedom. In the case that the number of these exceed
present bounds from big bang nucleosynthesis (BBN), the
requirement that these  not be in thermal equilibrium during BBN
gives a limit on $f$ of
approximately~\cite{Glob1,Glob2,Take,Hall,Gauge}
\begin{equation}
f \gtrsim 10 \, \mbox{keV}\,\, .
\label{flimit}
\end{equation}
A similar bound is obtained by demanding that SN cooling
not be modified in the presence of the above additional fields.

It is remarkable that this framework with a low NP scale,
$f\lesssim \Lambda_{\rm EWB}$ where $\Lambda_{\rm EWB}$ is the
electroweak symmetry breaking (EWB) scale, cannot be excluded by
direct experimental data.
In many late neutrino mass models, there are degrees of freedom
beyond  $\phi\,.$ These additional degrees of freedom can yield
indirect signals provided that standard cosmology is assumed. In
the case in which neutrino masses are protected by global or
approximate symmetries~\cite{Glob1} or the case with strong
dynamics (in which chiral symmetries are being broken by the
condensate)~\cite{Take}, light pseudo-Goldstone bosons (PGB) field are
typically present. Similarly, in models with gauge
symmetries~\cite{Gauge} the corresponding gauge boson masses are
suppressed, relative to $f$, by an additional gauge coupling $g,$
and therefore play a role similar to the one played by the PGBs.
This additional light fields interact with the plasma
through their coupling to the neutrinos. This happens even below the
BBN phase transition and may leave a trace in the observed cosmic
microwave background radiation (CMBR)~\cite{Glob1}.
\footnote{see also \cite{More} for related analysis.}

The focus of this work is to investigate other more direct ways of
testing the low scale models of neutrino masses.
We find that such
a possibility of a more direct probing of this class of models, at
present or in the not-too-distant future, does exist. The desired
signal would consist of a
dramatic modification of the supernova neutrino flux (diffuse or burst)
through interaction between the these neutrinos and the cosmic background
neutrinos (CBN). These interactions are mediated by the new  scalar
particles introduced by the NP.

In Section~\ref{Proc} we discuss the dominant processes which
modify the incoming SRN fluxes. We divide this section into two:
in~\ref{Res} we describe the resonant process which happens only
in a narrow range of parameter space, but leads to a spectacular
signal through what we denote as accumulative resonance;
in~\ref{Non-Res} we consider the  non-resonant processes which,
in conjunction with data from SN1987A,
yield a lower bound on $f$  comparable to the one from  BBN.  Also, at the
beginning of
Section~\ref{Proc} we summarize bounds on the model parameters
imposed by BBN. These bounds are sketched in Fig.\ref{figcompar}, along
with the parameter range for which the resonant process occurs.
Finally, we
conclude in Section~\ref{Conc}.

\section{Main Idea, Formalism and Important Processes}\label{Proc}

Our main idea in this work is to show that the presence of the
additional light bosons, required in the late neutrino masses
framework, can introduce a significant interaction between the SRN
and the CBN. This, in some region of the model parameter space,
can lead to a measurable modification of the incoming SRN flux.
The typical SRN energies are above average solar neutrino
energies and below the atmospheric ones. Consequently, this flux is
likely to be observed by SK~\cite{SK} and KamLAND~\cite{Kam}
in the near future, or by successor experiments. Thus, there is a
window (although not very wide) in which we can observe the
presence of both this extra light degrees of freedom and the CBN!
In this part we introduce the relevant part of the Lagrangian and
discuss important processes which yield the signal.

We first discuss resonant processes and present the phenomenon of
accumulative resonance which can yield a possible signal. Then we move
to discuss non-resonant processes, which, through the observation of
neutrinos from SN1987A yield a bound comparable with the
BBN one. Other implications of these processes, related to experiments
envisioned for the near future,
are also discussed.

Below EWB scale and close to the neutrino flavor symmetry breaking
scale the effective Lagrangian can be written as
\beq
{\cal L}_\nu^{\rm D}= {\cal L}_{\rm kin}+y_\nu \phi \nu
N+V(\phi)\,,\qquad {\cal L}_\nu^{\rm M}={\cal L}_{\rm kin}+y_\nu
\phi \nu \nu+V(\phi) \label{L}\,,
\eeq
where ${\cal L}_\nu^{\rm
D,M}$ stands for the Dirac and Majorana case respectively, ${\cal
L}_{\rm kin}$ denotes the kinetic part (for the gauge case
this contains interaction between $\phi$ and the additional gauge
bosons~\cite{Gauge}), $\nu$ represents an  active neutrino,
$V(\phi)$ is the scalar potential (for the global case
this contains interaction between $\phi$ and the additional
Goldstone bosons~\cite{Gauge}), and flavor and spinor indices are
suppressed for simplicity. The above implies that
\beq
m_\nu=y_\nu f\,.
\label{trirel}
\eeq
As we shall show below our signal is
similar in both the Dirac and Majorana cases. For simplicity through our
discussion below we omit the effect of neutrino mixings (apart from
the discussion related to the SRN flux).

 In order to establish a reference point, we first pause to summarize
the bounds on various models imposed by BBN constraints in terms of the Yukawa
couplings $y_\nu$. This will enable an evaluation of the feasibility of our
program.

1) The minimal model is of Majorana neutrinos with Abelian
symmetry. We  assume that the symmetry breaking scale, $f$, is
below the BBN temperature of about 1 MeV. Then during the BBN
epoch we cannot separate the Goldstone and the scalar (higgs) as
they are a single entity, a complex scalar field. The updated BBN
bound on the number of neutrinos is $N = 3.24 \pm 1.2$ at
95\%~\cite{PDG}.  The complex scalar adds 8/7 (neutrino) degrees
of freedom, so this additional degree of freedom can be
accommodated with the BBN bound above. However, there are other
cosmological bounds, such as SN cooling rate.
   If the BBN bound on relativistic degrees of
freedom should decrease by a significant amount, then the yukawa
$y_\nu$ would be subject to the upper bound obtained in the next
paragraph.

2) In the non-Abelian Majorana models, typically several complex
  scalars are present, which are not permitted to be  by BBN
considerations. Thus, in this case $y_\nu$  must be bounded from
above to ensure decoupling. This bound was derived in Ref.~\cite{Hall} and we
have also calculated this bound (as a check) by considering all
the processes that would produce G's. Recoupling via the $2\rightarrow 1$ process
$\nu\nu\rightarrow G$ takes place as the temperature falls to some value
$T_{\rm rec}$ determined by equating the decay rate at $T_{\rm rec}$ to the
Hubble expansion rate:
\begin{equation}
\frac{M_G}{3T_{\rm rec}}\ \frac{y_\nu^2\ M_G}{16\pi}\ = \
\sqrt{\frac{8\pi^5\ g } {45}}\ \frac{T_{\rm rec}^2} {M_{Pl}^2}\ \ ,
\label{rec}
\end{equation}
where $g$ is the number of degrees of freedom at $T_{\rm rec}.$
By requiring $T_{\rm rec} < T_{\rm BBN}$ we find
\begin{equation}
y_\nu \lsim 6 \times 10^{-7} (\rm{keV}/M_G)
\label{recbound}
\end{equation}

3) Finally, for the Dirac case, the absence of a negligible population of
right-handed (sterile) neutrinos $(N)$ in the bath disallows the reaction
$\nu N\rightarrow G$,
so that G's can only be produced via $\nu_L \nu_L \rightarrow G\ G$
(via $t$ channel $N$ exchange). Requiring that this process be out of equilibrium
at $T_{\rm BBN}$ yields  a BBN bound of
\begin{equation}
y_\nu \lsim 1 \times 10^{-5}\ \ .
\label{diracbound}
\end{equation}
The $s$-channel process  requires a chirality flip which makes the
bound weaker, as pointed out in Ref.~\cite{Glob1}. Note that this
bound in independent of the Goldstone mass.

\subsection{Resonance, Accumulative Resonance}\label{Res}
\begin{figure}[!hb]
\begin{center}
\includegraphics[height=3.cm, width=5cm]{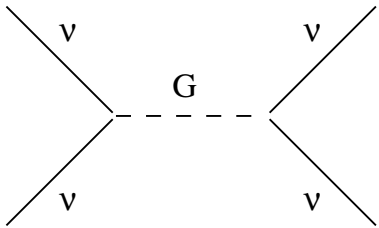}
\caption{Diagram representing resonant scattering. $G$ denotes a pseudo-Goldstone or a gauge boson.}
\label{figRes}
\end{center}
\end{figure}

The simplest possible process which will modify the spectrum is
the  resonant production of one of the above light bosons in the
collision of an SRN and a CBN. For simplicity we shall assume that
the boson couples only to neutrinos with a strength $y_\nu$ (this is a
good approximation for the case in which $y_\nu\lsim 10^{-6}$ as
discussed below). Thus
after being produced the boson, say a scalar, will decay back yet to a pair of
neutrinos where in our frame they have an energy spectrum flatly
distributed between 0 and the resonant lab energy $E^*$. The
diagram describing this process is shown in Fig.~\ref{figRes}.
\vspace*{-0.0cm}
For clarity, we frame the discussion which follows in terms of Dirac neutrinos.
Except for differing dynamics in the high temperature environment of the
supernova, the Majorana case is similar.

The resonant scattering affects the incoming SRN spectrum: the
energy of the incoming SRN  is now divided
between the two decay products, so that we expect to observe a
depletion in the expected spectrum for incoming neutrino with the
appropriate energies. In addition, since this process is
effectively 1$\to$2, one may expect  some depletion in the observed incoming neutrino spectrum
at appropriate energies.

Two questions are in order:

1. In view of the small upper limit on the interaction
strength (\ref{diracbound}) and the low density of the background neutrinos,
will the resonance process indeed produce the depletion discussed in the
preceding paragraph?

2. If the answer to (1) is affirmative,  can the
resultant depletion  be observed by
present or near future experiments?

Both these questions will be answered in the affirmative.
\subsubsection{Resonance: no cosmological expansion}

First, we consider the case of no expansion and begin by estimating the
mean free path  (m.f.p) $\lambda_{\rm Res}$ for the resonance process.

The cross section, written in Breit-Wigner form for the process in
Fig.~\ref{figRes} is roughly given by
\beq
\sigma_{\rm Res}\simeq
{y_\nu^4\over 16\pi} \,{ s\over \left(M_G^2-s\right)^2+M_G^2
\Gamma_\nu^2} \,,
\label{SigRes}
\eeq
where $G$ stands for
the gauge/Goldstone bosons
and $s$ is the square of the center of mass energy. In
addition $\Gamma_\nu$ is the decay width of the boson into
neutrino pair,
\beq
\Gamma_\nu\sim {y_\nu^{\;2}\, M_G\over
4\pi}\,.
\label{Gnu}
\eeq
Consider the case in which the SRN
energy, $E$, is on resonance, $E^*$,\footnote{We shall
neglect here the effect of thermal
broadening, effectively assuming that the background neutrinos are at
rest. We further discuss this point below. }
\beq
E^*\simeq{M_G^2\over 2m_\nu}\,,\label{Es}
\eeq
so that
\beq
\sigma_{\rm Res}\simeq {\pi\over M_G^2}
\,.\label{SigRes1}
\eeq
Consequently on the resonance\footnote{The result is qualitatively the same if
one averages over the width of the resonance.}
the m.f.p is give by
\beq
\lambda_{\rm Res}\approx {1\over n^0_\nu \sigma_{\rm Res}}\sim {M_G^2\over
  \pi n^0_{\nu}}\,,
\eeq
where $n_{\nu}= 3\pi \Gamma(3)\zeta(3)T_\nu^3\sim 56(1+z)^3 \,{\rm
  cm}^{-3}$, $n^0_\nu$ is the present background neutrino density
and $T_\nu$ is the background neutrino temperature with
$T^0_\nu\sim 1.6 \times10^{-4}{\rm \, eV}$ being the present one.
On carrying out the numerical substitution and using Eq.~\ref{Es}
we find\footnote{Note that for the lighter neutrino the m.f.p is
even somewhat shorter.}
\beq
\lambda_{\rm Res}\approx {1\over n^0_\nu \sigma_{\rm Res}}\sim {2 m_\nu E^*\over
  \pi n^0_\nu}\sim 5\times10^{-6}{\rm\, pc}\,{m_\nu\over 5\times 10^{-2} {\rm \, eV}}\,
\, {E^*\over 10 {\rm \, MeV}}\,.\label{opl}
\eeq
Since this is much smaller than a typical distance traveled by a
SRN neutrino, we find that the answer to Question 1. is positive for practically {\em any}
value of $y_\nu$. That
is, if a neutrino is produced with the appropriate resonance energy
then the process will go through.

However, this  analysis also implies  that the answer to the second
question is negative. The extent of the dip in the SRN  flux is
controlled by the width of the boson $G\,.$ This width is
tiny~(\ref{Gnu}), rendering it impossible at present or
in the near future to detect such a narrow depletion in the SRN
flux. The results are markedly different when cosmological expansion is
included, so that we now turn to consider this case.
\subsubsection{Resonance: case of cosmological expansion}

We start by finding the conditions on the coupling for which there is
sizable resonant
degradation of
of the {\em original} (not products of $G$-decay) flux  of supernova neutrinos.

The  the probability $P(E,z)$ that  a neutrino, created at red
shift $z$, with energy $(1+z)E$ arrives unscattered at the
detector with energy $E$ is given~\cite{Eberle:2004ua} by an
integration over proper time (converted to an integral over
intermediate red shifts $\bar z)$:
\beq
P(E,z)=\exp\left[-\int_0^z\;\frac{d\bar z}{H(\bar z)(1+\bar
    z)}\;{\bar{\,n}}_\nu\; \sigma_{\nu\nu\rightarrow \phi}(2m_\nu(1+\bar
z)E)\right]\ \ ,
\label{surv}
\eeq
where $\sigma$ is the resonant
scattering cross section, a function only of $s=2m_\nu(1+\bar
z)E,$ the (c.m. energy)$^2$ at the time of scattering, $\bar{\,n}_\nu$ is
the neutrino density (per flavor)  at redshift $\bar z$
and $H(\bar z)$ is the Hubble constant at redshift $\bar
z.$

For the purposes of this section, it is sufficient to employ a $\delta$-function
approximation for the cross section,
\beq
\sigma = \frac{\pi}{4}\ y_\nu^2\ \delta(s-M_G^2)\ \ ,
\eeq
so that the integral in (\ref{surv}) is trivially done, with the result
\begin{eqnarray}
P(E,z)& =& \exp\left[-\left(\frac{\pi y_\nu^2}{4 M_G^2}\ \right)\ n^0_\nu
\left(\frac{E^*}{E}\right)^3\  \left(\frac{1}{H(E^*/E)}\right)\right]\nonumber\\
&&{\rm for}\quad {E^*\over (1+z)} < E < E^*\ ,
\label{surv2}
\end{eqnarray}
and $P=1$ otherwise. Here  $E^*$ is the resonance energy $M_G^2/(2m_\nu).$
There will be large depletion of the initial SRN flux in this entire domain  if
\beq
\frac{\pi\ y_\nu^2}{ 4 M_G^2}\ \frac{n^0_\nu}{H_0} > 2\ \ ,
\eeq
which  gives (for $H_0 = 70 $
 km s$^{-1}$ Mpc$^{-1}$)
\beq
y_\nu\ \gsim\ 4\times 10^{-8}\ \ \frac{M_G}{1\ {\rm keV}}\ \ .
\eeq
In the Dirac  or Abelian Majorana cases, this permits a reasonable window of more than
two orders of magnitude in $y_\nu$ (see Eq.~(\ref{diracbound}))
for substantial depletion of the
SRN flux. For the non-abelian Majorana case, the window narrows to a bit more
than an  order of magnitude (Eq.~(\ref{recbound})). In this ``strong coupling''
regime we take $P=0$ in the
domain in Eq.~(\ref{surv2}).
Thus, with the appropriate constraint on $y_\nu,$ Question 1. is answered
in the affirmative: there will be substantial depletion of the original flux due
to resonant scattering. We now proceed to see how the cosmological evolution permits a
signal to be formulated for the resonant scattering.

\subsubsection{Accumulative resonance}

In the domain (\ref{surv2}), there will be resonant absorption out of the original
neutrino flux, but  some replenishment as well, from neutrinos
re-emitted in the decay of a $G$ produced in the domain in
Eq.~(\ref{surv2}).  More specifically, suppose that a neutrino
emitted with energy $\epsilon\ge E^*$ from a source at redshift $z$
undergoes resonant scattering at redshift $\bar z <z,$ so that
\beq
E^*=\epsilon\frac{1+ \bar z}{1+z}\ \ .
\label{reszzp}
\eeq
This
is followed by the emission of a decay neutrino with energy
$E^\prime = fE^*\, , 0\le f\le 1$ immediately following emission. The flatness
of the emitted-neutrino spectrum implies that that $f$ will vary uniformly over
the region [0,1].\footnote{We thank K. Hikasa for drawing our attention for this
issue.} In that
case the observed energy at the present era is
\begin{eqnarray}
E& = & \frac{fE^*}{1+\bar z}\nonumber \\
& = & \frac{f\epsilon}{1+z}\nonumber\\
& = & f E_{\rm unscattered}
\label{eobserv}
\end{eqnarray}
where $E_{\rm unscattered}\equiv \epsilon/(1+z)$ would be the observed energy of the neutrino
in the absence of resonant scattering. This shows how the entire allowed region of energy
below $E^*$ is populated by rescattered neutrinos, with the energies shifted downward from
the original spectrum. Especially interesting is the spectrum for $E_{\rm unscattered}$
at or just below the limit $E^*$: in that case, the only replenishment of flux is
from the tail of the decay distribution $(f\simeq 1)$, from resonant production that has
taken place only recently. (From Eq.~(\ref{reszzp}, on can see that the conditions for
this are that $\bar z \simeq 0.)$ The restriction to $f\simeq 1$ implies very little
replenishment, so that {\em a dip  at $E=E^*$ should be a universal
feature of the final spectra observed.} This is a qualitative response, in the affirmative,
to Question 2: the spectrum with absorption will show a dip at $E=E^*$, and will be shifted
downward from the spectrum absent resonant absorption. The complete effect of neutrinos
emitted with non-resonant energies, passing through resonance, and then replenishing
the flux at lower energies, is what we call accumulative resonance.

\subsubsection{A note on thermal broadening}
In the presence of the cosmological expansion, the effect of
thermal broadening (because the CBN are not at rest) on our
principal result (the universal dip described in the previous
subsection) is negligible. The argument is as follows: after
decoupling, the CBN spectrum is Fermi-Dirac, but in momentum
rather than energy, even into the non-relativistic region. Thus
the effect of thermal broadening is the introduction of a momentum
spread in the target neutrinos of ${\cal O}(T_\nu).$ The principal
feature of our result, the dip at neutrino energy $E^*$, occurs
with neutrinos undergoing resonant scattering in the recent era,
where the CBN is completely non-relativistic . In that case, the
fractional energy shift of the target neutrinos is ${\cal
O}(T_\nu^2/m_\nu^2)\sim 10^{-3}$ (unless the lightest neutrino has
mass $\lsim 10^{-4}$ eV). As can be seen from Eq.(\ref{surv}), the
effect of this uncertainty is the introduction of a spread of
${\cal O} (10^{-3})$ in the value of $\bar z,$ the red shift at
resonant scattering. The consequence, after integrating over red
shift of the source, is that the sharpness of the dip at $E^*$ in
the observed energy spectrum is softened by effects of ${\cal O}
(10^{-3})$ rather than  ${\cal O} (y_\nu^2) \sim 10^{-15}$,
corresponding to the intrinsic width of the resonance. Thus, even
with thermal broadening, the relative sharpness of the dip is
preserved.

\subsubsection{Event Rates}

\begin{figure}[!hb]
\begin{center}
\includegraphics[height=9.cm, width=13cm]{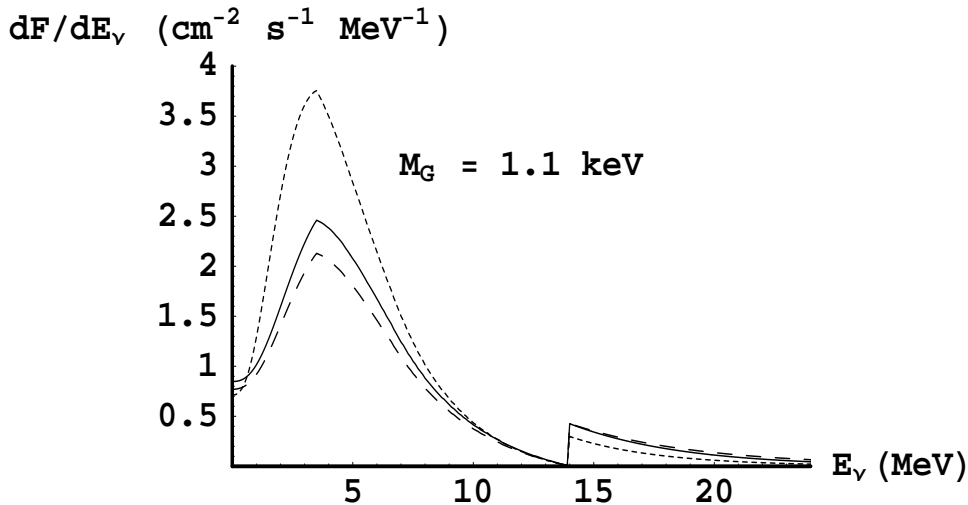}
\caption{Depleted incoming SRN  flux due to the resonance for
three different assumptions about the source flux. Dotted, solid
and dashed curves are fits designated as LL, TBP and KRJ,
respectively, after Eq.~(\ref{alphafit}).}
\label{figdFdEinterx}
\end{center}
\end{figure}

Next we consider the effect of the accumulative resonance on the total SRN
differential flux.
We first present the standard expressions for the SRN flux. Then
we shall discuss how to incorporate the accumulative resonance effect.
The differential flux of Supernova Relic Neutrinos (SRN) is
given by
\begin{equation}
\frac{dF}{dE} = \int_0^{zmax} R_{\rm SN}(z)\left<
\frac{dN(\epsilon)}{d\epsilon} \right>_{\epsilon = (1+z)E}\;(1+z)
\left|\frac{dt}{dz}\right| dz\; , \label{flux0}
\end{equation}
where for heuristic purposes we adopt as our standard the Fermi-Dirac
distribution
\begin{equation}
\frac{dN(\epsilon)}{d\epsilon}={\cal E} \times \frac{120}{7\pi^4}\times
\frac{\epsilon^2}{\left(T^{\rm SN}_\nu\right)^4}\times
\frac{1}{\exp\left(\frac{\epsilon}{T^{\rm SN}_\nu}
\right)+1} \; .
\label{fermidirac}
\end{equation}
The constant ${\cal E} = 0.5\times 10^{53} $  ergs is the total
energy carried by each flavor of neutrino.  The temperature for
the electron antineutrinos is $ T^{\rm SN}_{\bar\nu_e} = $ 5 MeV  and for the
non-electron
neutrinos and antineutrinos is $ T^{\rm SN}_{\nu_x} = $ 8 MeV. However, a more general form
\begin{equation}
 \frac{dN}{d\epsilon}=\frac{(1+\alpha)^{1+\alpha}{\cal E}}
  {\Gamma (1+\alpha)\bar\epsilon^2}
  \left(\frac{\epsilon}{\bar \epsilon}\right)^{\alpha}
  e^{-(1+\alpha)\epsilon/\bar\epsilon},
  \label{alphafit}
\end{equation}
has been proposed~\cite{Keil:2002in} which provides a good fit to
simulated explosions~\cite{Totani:1997vj} of high-mass progenitors. Here
where $\bar\epsilon$ is the average antineutrino energy at the source and the values of the fitting
parameters $\bar\epsilon$ and $\alpha$ for the $\bar\nu_{\rm{e}}$ and $\nu_{\rm {x}}$ spectra
from three different groups~\cite{Totani:1997vj, Thompson:2002mw, Keil:2002in}
(designated as LL, TBP, and KRJ, respectively),
summarized in Table 1
of Ref.~\cite{Ando:2004hc}. In Fig.~\ref{figdFdEinterx} we
will show the spread in our results obtained from the different spectra.

Since we are considering the case of Dirac neutrinos, we take the CBN to consist of
equal mixtures of left- and right-handed non-relativistic neutrinos and antineutrinos,
with total number
equal to two degrees of freedom in equilibrium during BBN. The resonant scattering will take
place between  right-handed SRN antineutrinos and CBN neutrinos, as
well as  left-handed SRN antineutrinos and CBN neutrinos. The problem becomes complex since
the spectrum of SRN neutrinos and antineutrinos are different. Again, for illustrative purposes,
we present results for the simplified case in which only SRN right-handed antineutrinos undergo
resonant scattering, but from a CBN left-handed neutrino population given by $n_0,$
as defined after Eq.~(\ref{surv}).

The Jacobian factor in Eq.~)(\ref{flux0}) is
given by
\begin{equation}
\frac{dt}{dz}=-\left[100\frac{\rm km}{\rm s~Mpc}\:h\:(1+z)
\sqrt{\Omega_M(1+z)^3\; + \Omega_\Lambda}\right]^{-1}\:
\end{equation}
with  $ \Omega_{M}=0.3 $, $ \Omega_\Lambda = 0.7 $ and $h$ = 0.7.
The (comoving) rate of supernova formation $R_{\rm SN}$ is parameterized as
\begin{equation}
R_{\rm SN}(z)=\left(\frac{0.013}{M_{\odot}}\right)\stackrel{.}{\rho_*}(z)
\end{equation}
where
\begin{equation}
\stackrel{.}{\rho_*}(z)=(1 - 2)\times10^{-2}M_{\odot}{\rm yr}^{-1}
{\rm Mpc}^{-3}\times(1+z)^{\beta}.
\end{equation}
and the exponent changes at $z=1$, from
$\beta \sim 2-4$ (for $0 < z <1$) to
$\beta \sim 0$ (for $z > 1$)
The uncertainty in parameters describing $R_{\rm SN}(z)$
comes from the uncertainty in
present knowledge of the Cosmic Star Formation Rate (CSFR)~\cite{Ste}.
In this paper we choose ``median'' values for these parameters~\cite{Ste},
${\rm R_{{\rm SN}}}(0) = 2 \times
10^{-4} \, {\rm yr}^{-1} \, {\rm Mpc}^{-3}$,
 $\beta = 2$ (for $0 < z <1$) and
$\beta = 0$ (for $z >1$).

The fact that the SN density is either constant (for far ones) or
increasing with the distance (for near ones) is of  great
importance. It implies that most of the incoming neutrinos
originate from distant SN (for far SN, the flux decreases like
square of the distance while their density grows like the cube of
the distance). Thus these neutrinos are redshifted and go through
the resonance. The detection process is sensitive only to the flux
of incoming anti-neutrino electrons. Consequently, to include
contributions to the flux from the muon/tau and electron neutrinos
we use the corresponding temperature in the expression for the
neutrino spectrum, and include a mixing factor which for the
electron neutrinos is 0.69 and for the muon neutrinos is 0.31.
As discussed in Refs.~\cite{Dighe:1999bi,
Ando:2002zj,Lunardini:2004bj}, the relation between the
$\bar\nu_e$ spectrum observed on Earth to the various neutrino
spectra at production depends critically on whether the neutrino
mass hierarchy is normal or inverted. If normal $(m_3>m_2>m_1)$,
then strong matter effects cause the $\bar\nu_e$ at production to
emerge from the stellar surface as the lightest eigenstate
$\bar\nu_1,$ with electron component $|U_{e1}|^2\simeq 0.69.$ The
small mixing of the electron with the third eigenstate
$|U_{e3}|^2\ll 1$ allows an equivalent two-flavor picture, with
the result that neutrinos produced in the supernova as
$\bar\nu_x,\; x=\mu\ \mbox{or}\ \tau$ will be received at Earth as
$\bar\nu_e$ with probability 0.31, and with energies corresponding
to the $\bar\nu_\mu$/$\bar\tau$ spectrum at production. For the
case of the inverted hierarchy, $\bar\nu_e$'s produced in the
supernova emerge as the lightest mass eigenstate , now
$\bar\nu_3$. For $\sin^22\theta_{13}\lsim 10^{-6}$ the resonance
is non-adiabatic and there is complete conversion
$\bar\nu_3\rightarrow \bar\nu_1$. This case then is the same as
for the normal hierarchy. The adiabatic case
($\sin^22\theta_{13}\gsim 10^{-4}$) is very different: the
original $\bar\nu_e$'s remain as $\bar\nu_3$ when emerging from
the stellar surface, contributing negligibly to the $\bar\nu_e$
flux at Earth. The entire $\bar\nu_e$ flux at Earth then
corresponds to the original $\bar\nu_x$ produced in the supernova.
For intermediate values of $\sin^22\theta_{13}$, the situation
is of course more complicated. In this paper we will consider only the
normal hierarchy. This can
be regarded as conservative, since in some cases the $\bar\nu_x$
spectrum (which generates the $\bar\nu_e$ signal in the adiabatic
inverted hierarchy scenario) is harder than the $\bar\nu_e$
spectrum, and will give rise to more events which escape the low
energy cuts.

In this context, we also note that  Earth matter effects have
been shown to modify the observed fluxes and spectra on
earth~\cite{Earth1}. Since however the hierarchy in average
energies between the $\bar \nu_e$ and the other flavor is milder
than thought this effect is expected to be
subdominant~\cite{Earth2}. Thus for simplicity and since these
effects are not expected to induce gross modification of the
observed spectrum we neglected them altogether.

In order to obtain the observed spectrum, we note again that all
neutrinos emitted from a source at redshift $z$ with energies
$\epsilon$ {\em outside} the window $E^*<\epsilon< (1+z)E^*$ will
arrive at $z=0$ without undergoing resonance, and with a flux
given by Eq.~(\ref{flux0}) with energies $E>E^*$ and $E<E^*/(1+z).$
For source energies in the resonant absorption region,  all of the
original flux will undergo resonance absorption followed by decay
into a flat spectrum. Since all energies, both before and after
absorption, are redshifted at the same rate, one can obtain the
rescattered spectrum by generating neutrino numbers according to
Eq.~(\ref{fermidirac}) in energy slices $\Delta\epsilon$ at the
source (for $\epsilon$ in the absorption region), and
redistributing this number according to Eq.~(\ref{eobserv})
uniformly over the observed energy region $0<E< \epsilon/(1+z).$

\begin{figure}[!hb]
\begin{center}
\includegraphics[height=9.cm, width=13cm]{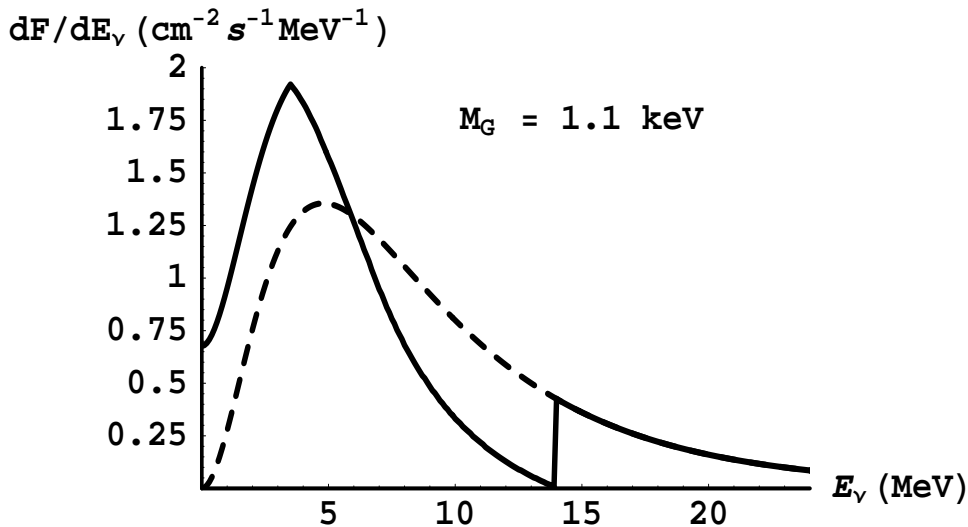}
\caption{Depletion in the incoming SRN  flux due to the
  resonance (solid curve)~\cite{Miami04},
compared to SRN flux without the
resonance (dashed curve). The source flux is Fermi-Dirac.}
\label{figdFdE}
\end{center}
\end{figure}

In Fig.~\ref{figdFdE}
we show the resulting  differential flux (with source flux given by Eq.~(\ref{fermidirac}),
with and without the
accumulative resonant effect, integrated over redshift up to
$z=4$,
 and for $M_G=1.1$ keV.
As discussed following Eq.~(\ref{eobserv}),
there is a sharp dip  at $E=E^*\equiv M_G^2/2m_\nu$ for all values of $z$.
To demonstrate the spread introduced through differing assumptions about the source flux, we
show in Fig.~\ref{figdFdEinterx} the flux with accumulative resonant effect for the three choices
discussed in the context of Eq.~(\ref{alphafit}).

\begin{figure}[!hb]
\begin{center}
\includegraphics[height=9.cm, width=13cm]{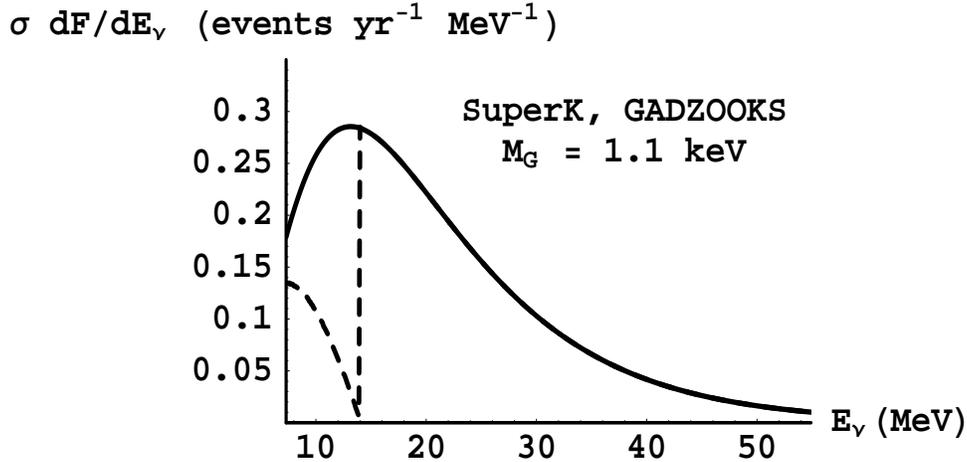}
\caption{Depletion in the incoming SRN  flux folded with the cross
section for detection with SK and GADZOOKS (dashed curve) compared
to no resonance case (solid curve). Source flux is Fermi-Dirac. The essential
features of the accumulation
resonance remain unmodified.}
\label{figdFdEC}
\end{center}
\end{figure}

To estimate the event rates for SK and GADZOOKS~\cite{GAD} (SK
enriched by Gd) we show in
Fig.~\ref{figdFdEC} the differential neutrino flux  folded with the
detection cross section.
This is for the inverse beta decay
induced by the anti-neutrino capture in the detector~\cite{InvB}.(For calculating
the event rate we have used the quasielastic
neutrino-nucleon cross section given in Ref.~\cite{Strumia:2003zx}.)
The shape of the differential rate is modified due
to the energy dependence of the cross section, which increases quadratically
with energy. The main features of the effect due to the accumulation
resonance such as the location of the dip and its width
remain unmodified.

\begin{figure}[!hb]
\begin{center}
\includegraphics[height=9.cm, width=13cm]{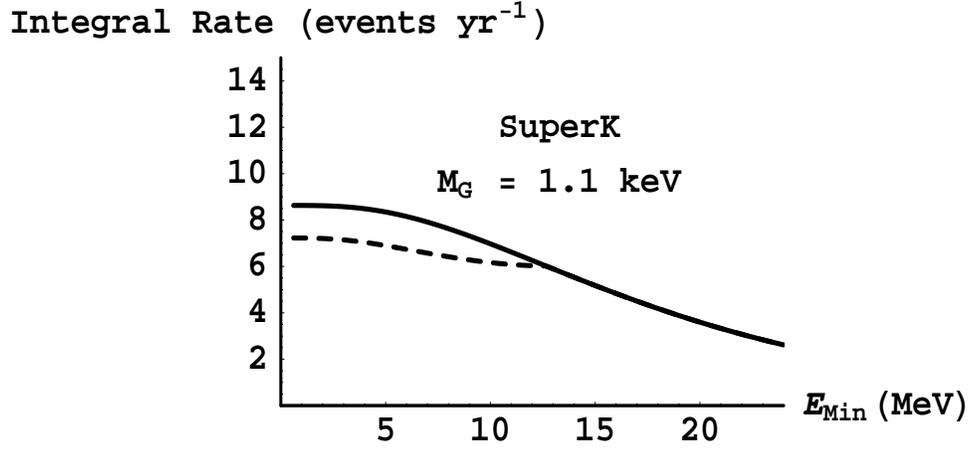}
\caption{Integrated event rates for GADZOOKS and SK. The threshold for GADZOOKS is
10 MeV and for SK
  it is 18 MeV. Source flux is Fermi-Dirac.}
\label{figSKF}
\end{center}
\end{figure}

\begin{figure}[!hb]
\begin{center}
\includegraphics[height=9.cm, width=13cm]{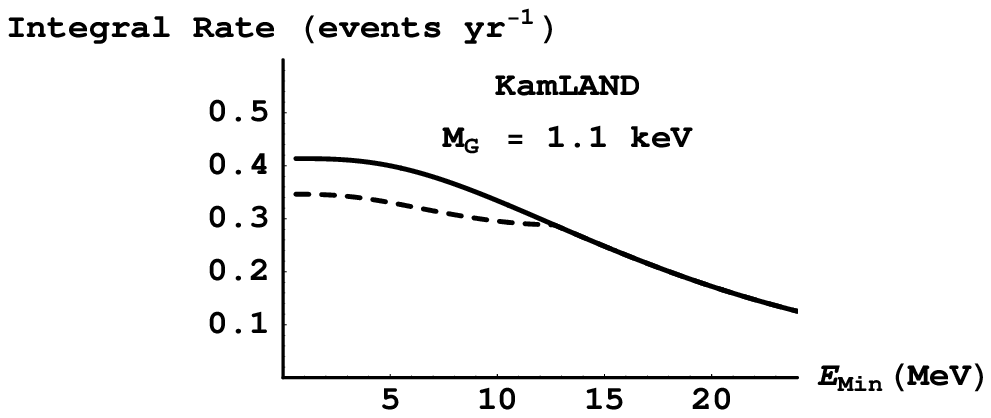}
\caption{Integrated event rates for KamLand. The threshold for KamLand
  is 6 MeV.}
\label{figKamF}
\end{center}
\end{figure}

The differential rate for SK and Kamland is rather low regardless of the
presence of the accumulative resonance. We therefore present in
Figs.~\ref{figSKF},\ref{figKamF} the integrated flux for SK and KamLand,
with and without accumulative resonance,
respectively. Since the energy thresholds for SK and KamLand are
18 and 6 MeV respectively, we note the interesting
feature that  the event rate for SK
is roughly unmodified (the 18 MeV threshold is well above the absorption
region) while a  suppression of roughly 25\%
is obtained for KamLand.
This feature gains definition in Fig.~\ref{figdFdEC}.
The GADZOOKS experiment has a much lower background (due to the
ability of identifying the emitted neutron) and therefore may be able
to provide a differential rate information.

\vspace*{-0.0cm}

\vspace*{-0.0cm}

\vspace*{-0.0cm}

\vspace*{-0.0cm}

\vspace*{-0.0cm}

\subsection{Non-resonance}\label{Non-Res}

The effectiveness of the resonance process in redistributing the $\rsn$ flux
requires the resonance energy to
be rather  close to the  peak energy $E_{\max}\simeq 3T^{\rm SN}_{\bar\nu}$ of the
$\rsn$ (perhaps redshifted by a factor of 2)  so that the boson mass
must be in the range of 1 keV,
\beq
M_{\rm boson}\simeq \sqrt{2m_\nu (1.5\, T^{\rm SN}_{\bar\nu})}\sim 1{\rm\,keV} \,
\left({E\over 10{\rm\,
    MeV}}\right)^{1/2}\,\left({m_\nu\over 0.05{\rm\,
    eV}}\right)^{1/2}\,.
\eeq
We note that the BBN constraint (\ref{flimit}) implies that
the mass of the symmetry breaking scalar $M_\phi\sim 10{\,\rm keV}$
is typically above the resonance mass (not that far though).
Thus it is more likely that the other light bosons (Goldstones or gauge) whose
masses are almost unconstrained are required to provide a resonant channel
for the $\rsn$ scattering.

\begin{figure}[!t]
\begin{center}
\includegraphics[height=6cm, width=9cm]{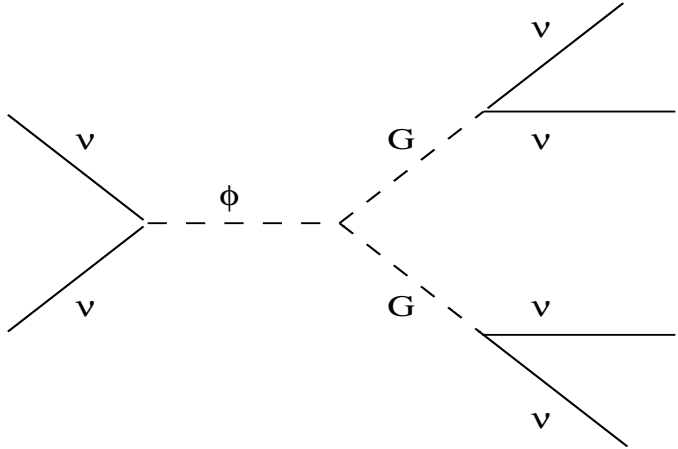}
\caption{Two Goldstones production through Higgs exchange, off-resonance process.
}\label{figOffRes}
\end{center}
\end{figure}

In view of this restriction, it is important to check whether other
non-resonant processes can become important.
One interesting possibility within the present dynamical framework
is  shown in
Fig.~\ref{figOffRes}.
Again, this presumes the existence of either
light Goldstone bosons~\cite{Glob1,Glob2,Take,Hall} or
light gauge bosons~\cite{Gauge}.
For $s\ll M_\phi^2\sim f^2$ we can estimate the
cross section to be
\beq
\sigma_{\rm GG}\sim{y_\nu^2 f^2\over m_\phi^4}
\sim {y^2_\nu\over f^2}\sim {y_\nu^4\over m_\nu^2}\,,\label{SGG}
\eeq
where we assume that the light bosons are produced on shell.  This
implies
\beq
M_G\leq \sqrt{2m_\nu E}/2
\label{Gshell}\,.
\eeq
To have a substantial scattering we require  the non-resonant mean free path
$\lambda_{\rm non-res}$ to be smaller than $H^{-1}$,
\beq
\lambda_{\rm non-res}H\simeq{H m_\nu^2
\over n_\nu y_\nu^4}\ll1
\,,
\eeq
which yields a lower bound on $y_\nu$
\beq
y_\nu > \left({H m_\nu^2
\over n_\nu}\right)^{1/4} \sim 10^{-6}\, \left({m_\nu\over 0.05{\rm\, eV}}\right)^{1/2}
\,.\label{lamp}
\eeq
This requirement is valid for any value of $M_G\,.$

The above process can have an effect on the SRN flux.
If $M_G<2m_\nu$ and there is sufficient optical depth,
all the SRN will be transformed into invisible Goldstones and
the signal is lost (for a related effect, see~\cite{Nussinov:1982wu}).
If $M_G>2m_\nu$ then the process can effectively be characterized as $\nu\to 4\nu,$
implying a substantial shifting of the entire SRN spectrum to lower energies.
For a point source at distance $\ell$, the condition analogous to Eq.~(\ref{lamp})
for sufficient optical depth is
\beq
y_\nu \geq 3.3\times 10^{-6} \, \left({3000 \,{\rm Mpc}\over \ell}\right)^{1/4}
\eeq
where $l$ is the distance travelled by the SRN.
For SN1987A, $\ell=50,000$ pc, the fact that non-resonant scattering have not
 occurred, i.e. neutrinos
with undegraded energy were observed ~\cite{sn1987a}, gives an independent upper
bound on $y_\nu$,
\beq
y_\nu\lsim 5.5
\times  10^{-5}\,.
\label{snbound}
\eeq
This bound is comparable to the cosmological one in
Eq.(\ref{flimit}). However, considerably more detailed work is
required in order to establish such a bound: a combined likelihood
analysis in the symmetry breaking scale and the parameters
describing the neutrino spectrum  needs to be done in order to
establish confidence levels for all variables~\cite{raffelt},
followed by marginalizing on the spectrum parameters. In the
meantime, we adopt (\ref{snbound}) as a rough, and preliminary,
indication that this bound can be comparable to others we have
mentioned.

\vspace*{-0.0cm}

\section{Conclusions}\label{Conc}

In the models in which neutrino masses are light due to the
low-energy symmetry breaking scale, extra light bosons are
typically present. These light bosons couple to neutrinos with the
coupling that
 is proportional
to their masses and therefore directly related to the symmetry
breaking scale.
We have shown that, in principle, one can measure this coupling
because SN neutrinos
interact with cosmic background neutrinos via these
bosons modifying the SN neutrino flux dramatically.

We have discussed two types of processes that are present due to these
interactions.  The first
is a low energy analog of a Z-burst~\cite{weilerfargion} where neutrinos interact
producing an on-shell boson
 which subsequently  decays to a pair of neutrinos. The expansion
of the universe allows for a wide range of energies in which such
a mini Z burst can occur. We characterize this process as
accumulative resonance. The second is a non-resonance process
which leads to a global degradation of energies in the supernova
neutrinos flux. The observation of neutrinos from the 1987a
supernova yields an important constraint on the parameter space of
the above models since the observations were fully consistent with
no such degradation.

As is often the case with observations related to neutrino physics,
the signal we find is currently beyond the limit
of each individual present experiment. The strength of our signal
can improve once data from several experiments are combined.
For example, our analysis reveals  that a robust prediction of
these models is that results from SK
will be unaffected by the above processes while KamLand should
observe a suppressed flux. However, in order to
be convinced that depletion is observed it is desirable
to  actually observe the predicted  dip in the flux, which requires certain amount of
information on the energy dependence of the flux.
This can be obtained in
the future via water Cerenkov detectors enriched with added Gadolinium.
(the GADZOOKS~\cite{GAD} proposal).
This can be implemented in the very near future using an upgrade of SK, or
attained via future experiments such as HyperK, UNO etc...
Such and experiments can collect tens of
relic supernova neutrinos per year and provide us with information on
the energy dependence
of SN flux.
Furthermore they are more sensitive to observation of a single SN
event.
In principle these Megaton scale water Cherenkov detectors might
detect neutrino burst from {\cal O} (Mpc)
distant SN with average rate of one burst per two years or so (see
{\it e.g.}~\cite{ABY} and references therein).
In this case there are two possible scenarios.
The first, less exciting, is that the future
 burst of SN neutrinos
will have differential flux
consistent with the one observed from SN1987A.
In this case we expect the bound on the model parameters to be improved.
Alternatively, a more exciting possibility is that the incoming neutrinos are
absent, or are severely degraded in energy relative to expectations.
This can then be interpreted as
a measurement of the coupling between the neutrino, the Higgs and
the Goldstone bosons which yielded the degradation in the flux through
the non-resonance process.

We note parenthetically that  we have for simplicity
neglected modification of the neutrino spectra due to shock wave
effects~\cite{shock1}. These effects can change the spectral
features, inducing non-adiabatic transitions, in a manner that
depends on the neutrino flavor parameters~\cite{shock2}. However,
since the exact dynamics related to the shock propagation in the
SN is not well understood and since the above effects are
red-shift dependent and will be smeared out once the integration
over $z$ is applied we shall not include then in our computations.
In this context it is important to emphasize that the position of
the universal dip will not vary with redshift.

It is also important to stress that in this work we focus on
the modification of the dynamics of the SN neutrinos while they
propagate in space outside the SN. Interesting effects may be
induced by the presence of the new light degrees of freedom inside
the SN~\cite{Farzan}. These, however, are model dependent; for
example, in the case of late Dirac neutrino masses the overall
effect is expected to be miniscule~\cite{Glob1}. The Majorana case
is more involved since the extra bosons are expected to be
thermalized inside the SN core~\cite{Glob1,Farzan} through their
reactions $\nu\nu\leftrightarrow G.$ Consequently, the dynamics
inside the SN might be significantly modified due to the presence
of new flavor and lepton violating interactions. This requires a
more detailed study which is beyond the scope of our paper. Note
that it is likely that only the (pseudo) Goldstone boson will be
light enough to be thermally produced inside the
core~\cite{Glob1,HMP}. Thus the modification of the the neutrino
spectra~\cite{DH} is expected to be below the 10\% level expected
just by counting degrees of freedom, which is probably smaller
than other systematic effects which have not been included in this
analysis.

\begin{figure}[!hb]
\begin{center}\hspace*{-.1cm}
\includegraphics[height=8.5cm, width=13.5cm]{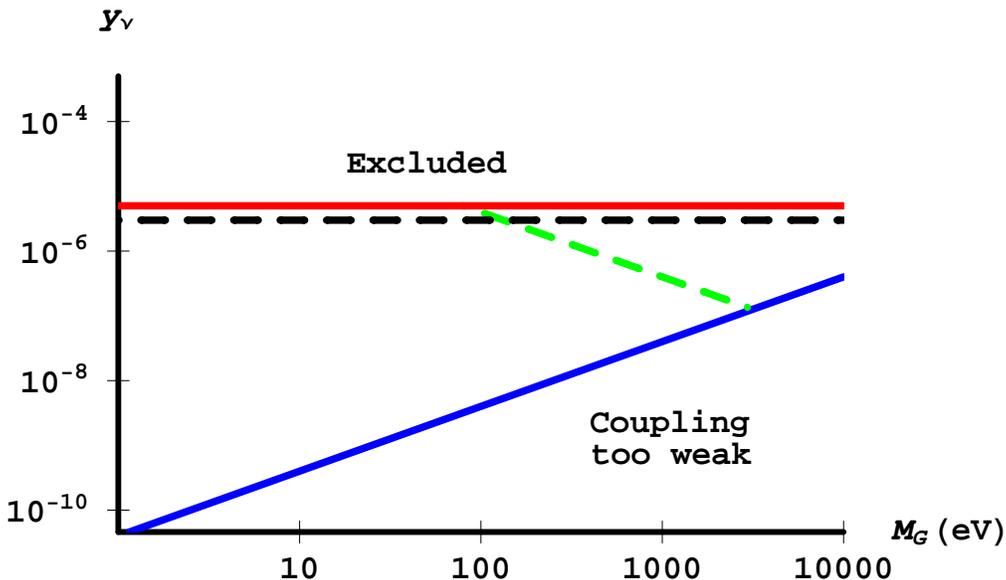}
\caption{The cosmological bounds and
the regions for the supernova neutrino spectrum distortion due to the resonance
and non-resonance processes for a single Majorana (Dirac)
neutrino for an abelian (non abelian) model
 are shown in ($y_\nu,M_G$) plane.
The region above the red (solid horizontal) line
is excluded by the BBN constraint (for the Dirac case), SN cooling (for Majorana case)
 and due to the observation of (undegraded) SN1987A neutrinos.  In the region below
the blue (solid slanting) line the mean free path is too long for
the resonance to occur. The region above
the green (dashed slanting) line, which is relevant only for
 the non abelian Majorana case, is the region excluded by the BBN constraint.
 The region above the black (dashed horizontal) line is the region of the
 future experimental sensitivity to the
 observation/non-observation of the SRN neutrinos due to the the non-resonant processes.
}
\label{figcompar}
\end{center}
\end{figure}

 A summary of the available parameter space, subject to BBN bounds, for
observation of the resonant process is given in Fig.~\ref{figcompar}.
It is clear that BBN constraints are most constrictive in the case of
Majorana neutrinos with a non-Abelian symmetry breaking sector. Least
restrictive is the Dirac, Abelian, scenario; it also has the least impact on
SN dynamics at the source.


 We stress that observation of the above signals may not only shed
light on the origin of neutrino masses but also yield
an indirect observation of the elusive background relic neutrinos.
In addition, we note that the above processes may be induced by other
light degrees of freedom which couples to neutrinos. Consequently
our signal may provide a test for other frameworks apart from the late
neutrino masses one.

\vspace*{.2cm}
{\bf Acknowledgements}

The authors thank the Aspen Center for Physics where
this work was initiated.
We wish to thank A. Burrows, G.D. Kribs, G. Steigman, H. Davoudiasl,
H. Murayama, J.G. Learned, K. Hikasa, L.J. Hall,
R. Kitano, S. Oliver for helpful discussions.  IS thanks J. Baker for
the help with the numerical evaluations. This research was supported in
part by the National Science Foundation Grant No. PHY99-0794.
HG is supported by NSF under grant
PHY-0244507,
GP is supported by DOE under contract DE-AC03-76SF00098 and IS is
supported
by DOE under contracts
DE-FG02-04ER41319 and
 DE-FG02-04ER41298 (Task C).  IS would like to thank KITP Santa Barbara and
LBNL Theory Group for
hospitality while this work was being completed.

\end{document}